\documentclass[prb,twocolumn,aps,showpacs,superscriptaddress,amsfonts,amssymb,amsmath]{revtex4}
\usepackage{amsmath}
\usepackage{graphicx}
\begin{document}
\title{A possible minimum toy model with negative differential capacitance for self-sustained current
oscillation}
\author{Gang Xiong}
\affiliation{Physics Department, Beijing Normal University,
Beijing 100875, P. R. China}
\author{Z. Z. Sun}
\affiliation{Physics Department, The Hong Kong
 University of Science and Technology, Clear Water Bay,
Hong Kong SAR, China}
\author{X. R. Wang} \affiliation{Physics
Department, The Hong Kong
University of Science and Technology, Clear Water Bay,
Hong Kong SAR, China}

\date{Draft on \today}

\begin{abstract}
We generalize a simple model for superlattices to include the
effect of differential capacitance. It is shown that the model
always has a stable steady-state solution (SSS) if all
differential capacitances are positive. On the other hand, when
negative differential capacitance is included, the model can have
no stable SSS and be in a self-sustained current oscillation
behavior. Therefore, we find a possible minimum toy model with
both negative differential resistance and negative differential
capacitance which can include the phenomena of both self-sustained
current oscillation and $I-V$ oscillation of stable SSSs.
\end{abstract}

 \maketitle

\section{Introduction}
Following the early pioneering study\cite{chang,buttiker} on
vertical electron transport in superlattices (SLs), one of the
recent surprising discoveries is self-sustained current
oscillations (SSCOs) under a dc bias\cite{kwok,jwang}. A large
number of experimental and theoretical studies have focused on
different aspects of these oscillations. Experimentally, it is
known that SSCOs can be induced by varying the doping
density\cite{kwok}, temperature and magnetic field\cite{jwang}.
Theoretically, it is understood that SSCOs are accompanied by the
motion of boundaries of electric field domains
(EFDs)\cite{kastrup}. A discrete drift (DD) model capable of
describing both the formation of stationary EFDs and SSCOs emerged
after many tedious analysises and numerical
calculations\cite{bonilla,bonilla1}. Our understanding of SSCOs
was greatly advanced through numerical investigations of this
model\cite{bonilla2}. However, this nonlinear dynamical model is
too difficult to be solved analytically in a way that one can
understand its different types of attractor solutions and their
bifurcations. Thus, a minimum model with certain solvability may
give some help for understanding these phenomena.

Recently Wang and Niu(WN)\cite{wn} proposed a simple model with
many nice features. This model is capable of revealing
analytically the source of instabilities in the electron
transport. However, we can show that this model always has at
least one stable steady-state solution (SSS) under a dc bias. Thus
it does not support a SSCO, and it cannot be a minimum model for
the SSCO phenomenon. The question that we want to address is
whether it is possible to find a minimum model which is simpler
than the DD model. We shall show that a generalized WN model, by
including the negative differential capacitance, may have no
stable SSS, thus it can be a possible minimum toy model for SSCO
in superlattices. This paper is organized in the following way. In
Sec.\ref{WN}, we shall show that there is always a stable SSS in
the original WN's model no matter what kind of $I-V$ curve is used
for each barrier. Then we will derive a generalized WN model by
including negative differential capacitance in Sec.\ref{GWN}. In
Sec.\ref{NDC}, we shall show that this toy model with both
negative differential capacitance and NDR can include both $I-V$
oscillations of stable SSSs and SSCO behaviors if the model
parameters are properly selected.

\section{Existence of a stable steady-state solution in the original WN model}
\label{WN} Let us first briefly review WN's model. WN employed the
following three assumptions for their model of a superlattice of N
quantum wells. (1) Inside each quantum well, the electronic states
are described fully by time-independent quantum mechanics. (2)
Charge carriers are in local equilibrium with each well so that a
chemical potential can be defined locally. The chemical potential
difference between two adjacent wells is call the bias V between
them. (3) For a given bias V between two adjacent wells, a current
I(V) passes through the barrier between them. In the following,
$I_{i}(t)$ and $V_{i}(t)$ denote the electric current and bias
through the $i-$th barrier, $n_{i}(t)$ the {\it free} charge in
the $i-$th well. The current $I_{i}(t)$ is supposed to depend on
the bias $V_{i}(t)$ only. WN thus obtain the following dynamical
equations
\begin{equation}
    \frac{dn_{i}(t)}{dt}=I_{i-1}(t)-I_{i}(t),
    \label{wn01}
\end{equation}
\begin{equation}
    V_{i}(t)-V_{i-1}(t)=kn_{i}(t),
    \label{wn02}
\end{equation}
\begin{equation}
    \sum_{i}V_{i}(t)=U,
    \label{wn03}
\end{equation}
where
\begin{equation}
    k=\frac{4\pi}{\epsilon S}.
\end{equation}
\begin{figure}[t]
\begin{center}
  \includegraphics[height=7cm,width=8cm]{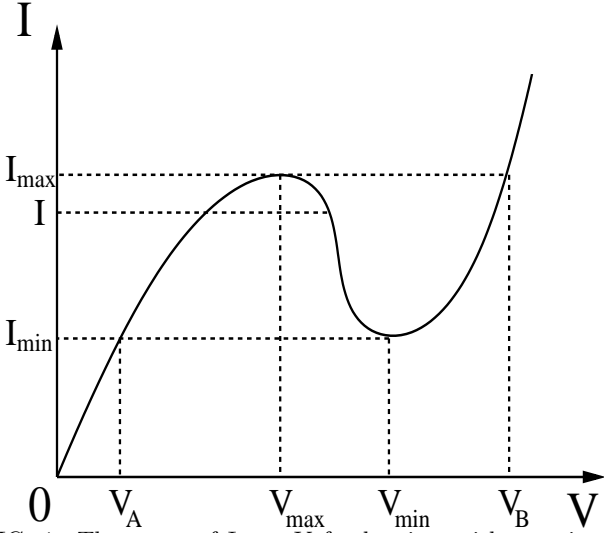}
\end{center}
\vskip -0.8cm
 \caption{The curve of I vs. V for barriers with a
region of negative differential resistance in the case that both
the minimum current and maximum current of all barriers are the
same.} \label{I_V1} \vskip -0.8cm
\end{figure}
Eq.(\ref{wn01}) and Eq.(\ref{wn02}) are the charge conservation
law and the Poisson equation for the $i-th$ well, while
Eq.(\ref{wn03}) is the constraint condition that the total bias
should be constant. As a first step, let us consider a simple case
that I-V curves of all barriers are the same, as shown in
Fig.\ref{I_V1}. Denote $(I_{max},V_{max})$ and $(I_{min},V_{min})$
as the maximum and minimum currents and the corresponding biases
in the I-V curve as shown in Fig.\ref{I_V1}. An SSS of
Eq.(\ref{wn01})---(\ref{wn03})
\begin{equation}
    V_{i}(t)\equiv V_{i}^{0}
\end{equation}
should satisfy
\begin{equation}
    I(V_{i}^{0})\equiv I
\end{equation}
where $I$ is the equilibrium current of the SSS. WN have proven
the following conclusion\cite{wn}.

{\bf Lemma 1.} {\bf (a)} An SSS is stable if all barriers at the
SSS are in their PDR regions, i.e.,
\begin{equation}
    \frac{dI_{i}}{dV_{i}}\Big|_{V_{i}=V_{i}^{0}}\equiv I^{\prime}_{i}(V_{i}^{0})>0;
    \label{wn04}
\end{equation}
{\bf (b)} if one and only one barrier of an SSS is in its NDR
region, then the SSS is stable if and only if
\begin{equation}
   \sum_{i=1}^{N}[I^{\prime}_{i}(V_{i}^{0})]^{-1}>0;
   \label{wn05}
\end{equation}
{\bf (c)} if two and more barriers are in their NDR regions, then
the SSS is unstable.

With the use of {\bf Lemma 1}, we can prove

{\bf Claim 1.} For Eq.(\ref{wn01})---(\ref{wn03}), there is at
least one stable SSS for each value of $U$ when all $I-V$ curves
are the same and has only one NDR region.

{\bf Proof.} When $U\in[0,NV_{max}]$ or $U\in[NV_{min},\infty)$,
an SSS with $V_{i}\equiv U/N$ for all barriers is stable according
to {\bf Lemma 1(a)} since $V_{i}=U/N$ drops in either lower-bias
PDR region $[0,V_{max})$ or upper-bias PDR region
$(V_{max},+\infty)$. Thus our proof can be restricted to the
region of $U\in(NV_{max},NV_{min})$, and all we need to do is to
find a stable SSS for each value of $U\in(NV_{max},NV_{min})$. Let
us start by considering an SSS with the first barrier in its NDR
region and all other barriers in their lower-bias PDR regions as
shown in Fig.\ref{I_V1}. The equilibrium current $I$ of this SSS
is slightly smaller than $I_{max}$, i.e., $\delta I=I-I_{max}<0$
is a negative infinitely small value. The bias on the $i-$th
barrier for this SSS is thus
\begin{eqnarray}
   V_{i}(\delta I)=V_{max}-\delta V_{i}
\label{claim01}
\end{eqnarray}
with
\begin{eqnarray}
    \frac{\delta V_{i}}{\delta I}=[I^{\prime}_{i}(V_{i})]^{-1}.
\label{claim02}
\end{eqnarray}
Then we have
\begin{equation}
  NV_{max}-\sum_{i=1}^{N}V_{i}(\delta I)=\delta
  I\sum_{i=1}^{N}[I^{\prime}_{i}(V_{i})]^{-1}.
\label{claim03}
\end{equation}
According to {\bf Lemma 1(b)}, this SSS is unstable when and only
when
\begin{equation}
    \sum_{i=1}^{N}[I^{\prime}_{i}(V_{i})]^{-1}\le0
\label{claim04}
\end{equation}
which leads to
\begin{equation}
  NV_{max}-\sum_{i=1}^{N}V_{i}(\delta I)=\delta I\sum_{i=1}^{N}[I^{\prime}_{i}(V_{i})]^{-1}\ge0.
\label{claim05}
\end{equation}
Let us denote the total bias of the above SSS with the equilibrium
current $I$ as $U(I=I_{max}-\delta
I)\equiv\sum_{i=1}^{N}V_{i}(I)$. Now let us look for a stable SSS
in the case when the total bias $U=U(I)$. If the above SSS itself
is stable, then it can be the SSS we look for. On the contrary, if
the above SSS is unstable, then by Eq.(\ref{claim05}) we have
$U(I)=\sum_{i=1}^{N}{V_{i}(I)}\le NV_{max}$ which means that
$U(I)$ is in the region $[0,NV_{max}]$. Thus there is also a
stable SSS in this case since we have shown that there is always
one stable SSS when $U\in[0,NV_{max}]$. Therefore, there is always
a stable SSS when the total bias
$U=U(I)=\sum_{i=1}^{N}{V_{i}(I)}$, regardless of whether the above
SSS is stable or not. Since $\delta I$ is regarded as infinitely
small, by Eq.(\ref{claim03}) the value $U(I)-NV_{max}$ is also
infinitely small. Thus we can say that we have proven the
existence of a stable SSS when $U\in[0,U(I)]$. Now let us further
decrease $I$ by $\delta I$ and consider the corresponding SSS of
the total bias $U=U(I_max-2\delta I)$ with only the first barrier
in the NDR region. Similar to
Eqs.(\ref{claim01})--(\ref{claim05}), we can obtain that this SSS
in unstable when and only when $U(I_{max}-2\delta I)\le
U(I_{max}-\delta I)$. Therefore, we can prove the existence of a
stable SSS when $U\in[0,U(I_{max}-2\delta I)]$. By decreasing $I$
step by step, we can go on with this procedure before $I=I_{min}$.
When $I=I_{min}$, the SSS we consider is as follows : the bias of
the first barrier is $V_1=V_{min}$ and the biases of all other
barriers are $V_{i}=V_{A}$. Thus we have proven the existence of a
stable SSS for $U\in[0,V_A+(N-1)V_{min}]$. Then, let us consider
the following series of SSSs and increase the equilibrium current
$I$ from $I_{min}$ to $I_{max}$ : the first barrier in a part of
its upper PDR region, i.e., $V_1\in(V_{min},V_B)$ and all other
barriers in a part of their lower PDR regions, i.e.,
$V_i\in(V_{A},V_{max})$. It is obvious that this series of SSSs
are stable because all barriers are in PDR regions. Thus we have
proven the existence of a stable SSS for
$U\in[0,V_{B}+(N-1)V_{max}]$ when $I$ is increased to $I_{max}$.
At this point, the first barrier is in its upper-bias PDR region
while all other barriers are at the cross point between the
lower-bias PDR region and the NDR region. To go further, we can
decrease $I$ and keep the first barrier in its upper-bias PDR
region, let the second barrier in its NDR region and all other N-2
barriers in their lower-bias PDR regions. Following the above
procedure, by decreasing $I$ step by step we can arrive at an SSS
that the biases of the first and second barrier are
$V_1=V_2=V_{min}$ while all other N-2 barriers are at $V_i=V_A$.
Thus we have proven the existence of a stable SSS for
$U\in[0,2V_A+(N-2)V_{min}]$. Then, follow the above procedure and
consider the following series of SSSs and increase the equilibrium
current $I$ from $I_{min}$ to $I_{max}$ : $V_1\in(V_{min},V_B)$
and $V_2\in(V_{min},V_B)$ while $V_i\in(V_{A},V_{max})$ for all
other barriers. Thus we have proven the existence of a stable SSS
for $U\in[0,2V_{B}+(N-2)V_{max}]$ when $I$ is increased to
$I_{max}$. Then, we decrease $I$, and keep the first and second
barriers in their upper-bias PDR regions, let the third barrier in
its NDR region and all other N-3 barriers in their lower-bias PDR
regions. Following the above procedure, we can prove the existence
of s stable SSS for $U\in[0,3V_B+(N-3)V_{max}]$ and so on so
forth. Finally, we can prove the existence of a stable SSS for
$U\in[0,NV_{max}]$. {\bf QED}.

\begin{figure}[b]
\vskip -0.8cm
\begin{center}
  \includegraphics[height=6cm,width=8cm]{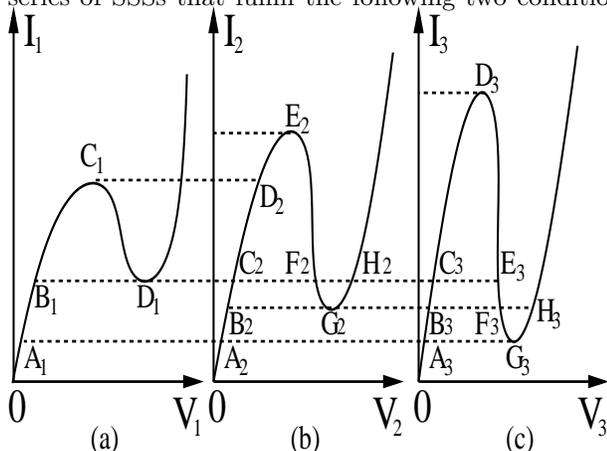}
\end{center}
\vskip -0.8cm
 \caption{The curve of I vs. V for (a) the first
barrier, (b) the second barrier, and (c) the third barrier. The
minimum current and maximum currents of them are not same.}
\label{I_V2}
\end{figure}

Summarize the above proof, we can find the following general
method to prove the existence of a stable SSS. Suppose that for
some $U_0>0$ we have proven the conclusion that there is always a
stable SSS for $U\in[0,U_0]$. Then, in order to extend the
conclusion to a larger region $U\in[0,U_1]$ with $U_1>U_0$, all we
need to do is to find a series of SSSs that fulfill the following
two conditions : (1) the maximum number of barriers in NDR regions
for each SSS is one; (2) the total bias of the series of SSSs
covers the region between $U_0$ and $U_1$. Using this method, we
can extend the conclusion of {\bf Claim 1} to a general case and
obtain

{\bf Claim 2.} For Eq.(\ref{wn01})---(\ref{wn03}), there is alway
one stable SSS for each value of $U$ when $I-V$ curves of barriers
are not necessarily the same while each barrier has only one NDR
region.

{\bf Proof.} Since $I-V$ curves of barriers are not necessarily
the same, let us denote $(I_{max,i},V_{max,i})$ and
$(I_{min,i},V_{min,i})$ as the maximum and minimum currents and
the corresponding biases in the I-V curve of the $i-$th barrier.
Without loss of generality, let us suppose that $I_{max,1}\le
I_{max,2}\le \cdot\cdot\cdot \le I_{max,N}$ as shown in
Fig.\ref{I_V2}. Let us start by considering a series of SSSs with
the first barrier lying between the origin and $C_1$ while all
other barriers lying in their low-bias PDR regions. Let us
increase the equilibrium current of these SSSs $I$ continuously
from zero to $I_{max,1}$. This series of SSSs are stable because
all barriers are in PDR regions. Now we arrive at an SSS with the
first barrier at $C_1$ while all other barriers in their low-bias
PDR regions. Let us denote the total bias of this SSS as $U_0$.
Then, we have proven that there is always a stable SSS for
$U\in[0,U_0]$. The rest task of the proof is to find a series of
SSSs that fulfill : (1) the maximum number of barriers in NDR
regions for each SSS is one; (2) the total bias of the series of
SSSs covers the region $(U_0,+\infty)$. Following the method in
{\bf Claim 1}, let us decrease $I$ from $I_{max,1}$ to $I_{min,1}$
and consider a series of SSSs with the first barrier lying between
$C_1$ and $D_1$ and the second lying between $D_2$ and $C_2$ while
all other barriers lying in their low-bias PDR regions. Then we
arrive at an SSS with the first barrier at $D_1$ and the second
barrier at $C_2$ while all other barriers lying in their low-bias
PDR regions. Let us denote the total bias of this SSS as $U_1$.
Thus we have proven that there is always a stable SSS for
$U\in[0,U_1]$. Then, keep the first barrier in its high-bias PDR
region, let $I$ increase, and move the second barrier along the
trace $C_2\rightarrow D_2\rightarrow E_2\rightarrow F_2$ and other
barriers correspondingly. If $I_{min,2}\ge I_{min,1}$, then the
second barrier can cross $G_2$ when the first barrier is still in
its high-bias PDR region. If $I_{min,2}<I_{min,1}$, then the
second barrier reaches $F_2$ when the first barrier reaches $D_1$.
At this point, the second barrier can not move continuously from
$F_2$ to $H_2$ with the above method applicable. In order to fill
the sudden jump from $F_2$ to $H_2$, we can do as follows : move
the first barrier from $D_1$ to $B_1$ while all other barriers are
unmoved. This decreases the total bias. According to the above
proof, there is a stable SSS for this total bias. Then, we can
first move the second barrier from $F_2$ to $H_2$ while keep all
other barriers in their low-bias PDR regions, then move the first
barrier along $B_1\rightarrow C_1\rightarrow D_1$ while keep the
second barrier in its high-bias PDR region and other barriers in
their low-bias PDR regions. By doing this, we let the second
barrier cross its minimum current point while keeps the number of
NDR barriers in this series of SSSs less than or equal to one.

Now let us consider the third barrier. If $I_{min,3}$ is larger
than $I_{min,1}$ and $I_{min,2}$, then it is easy to let the third
barrier cross its minimum current point. If $I_{min,3}$ is less
than one of $I_{min,1}$ and $I_{min,2}$, say,$I_{min,3}<I_{min,1}$
while $I_{min,3}>I_{min,2}$ as shown in Fig.\ref{I_V2}, then we
can use the method in the proceeding paragraph to let the third
barrier cross its minimum current point. If $I_{min,3}$ is less
than both $I_{min,1}$ and $I_{min,2}$, then we can do as follows.
When the third barrier reaches $E_3$, move the first barrier from
$D_1$ to $B_1$. Then, move the second barrier from $G_2$ to $B_2$
when the third barrier reaches $F_3$. After that, let the third
barrier cross its minimum current point while keeping all other
barriers in their low-bias PDR regions. Then let the first barrier
cross its minimum current point while keep the third barrier in
its high-bias PDR region and the second barrier in its low-bias
PDR region. When this is finished, let the second barrier cross
its minimum barrier with the use of the above method. Thus the
first, second and third barriers cross their minimum current
point. Keep this go on, we can cross the NDR regions of all N
barriers. This finishes the proof because the conclusion is
obvious when all NDR regions are crossed. {\bf QED}.

In the above proof, we restrict to the case that there is only one
NDR region in $I-V$ curve of each barrier. It is straightforward
to generalize the above proof for general $I-V$ curves that may
have a lot of NDR regions separated by PDR regions. In fact, one
can show that the above method works if the following condition is
fulfilled : each barrier always lies in PDR region when the
absolute value of its bias is above some upper-limit value. This
condition is generally fulfilled. Therefore, we can come to the
conclusion that there is always a stable SSS in WN's original
model no matter what kind of $I-V$ curve is used for each barrier.

\section{A generalized model with differential capacitance}
\label{GWN}

In this section, we shall derive a generalized WN model and
introduce negative differential capacitance. In the generalized
model, we consider each barrier as a narrow homogeneous dielectric
material. Then, instead of using the Poisson equation, we make use
of the Gauss' law for dielectric materials
\begin{equation}
    [D_{i}(t)-D_{i-1}(t)]S=4\pi n_{i}(t)
    \label{gwn01}
\end{equation}
where $D_{i}(t)$ the electric displacement in the $i-$th barrier
and $S$ the transverse area of the superlattice. We assume that
$D_{i}(t)$ only depends on the average electric field
$E_{i}(t)=V_{i}(t)/L_{i}$, $L_{i}$ the thickness of the $i-$th
barrier. Eq.(\ref{wn01}) still holds since it is the charge
conservation law. Eliminate $n_{i}(t)$ by Eq.(\ref{wn01}) and
Eq.(\ref{gwn01}), we have
\begin{equation}
    C_{i}(V_{i})\frac{dV_{i}}{dt}
    -C_{i-1}(V_{i-1})\frac{dV_{i-1}}{dt}
    =I_{i-1}(V_{i-1})-I_{i}(V_{i})
\label{gwn02}
\end{equation}
and
\begin{equation}
    \sum_{i=1}^{N}\frac{dV_{i}(t)}{dt}=0
\label{gwn03}
\end{equation}
where
\begin{equation}
    C_{i}(V_{i})=\frac{S}{4\pi}\frac{dD_{i}}{dV_{i}}
    =\frac{S}{4\pi L_{i}}\frac{dD_{i}}{dE_{i}}.
\label{gwn04}
\end{equation}
One can see that $C_{i}(V_{i})$ serves as the differential
capacitance of the $i-$th barrier because $dD_{i}/dE_{i}$ can be
viewed as the differential dielectric function of it. When the
conventional relation $dD_{i}/dE_{i}=\epsilon$ is substituted into
Eq.(\ref{gwn04}), one returns to WN's original model. Thus the
above model is a generalized version of WN's model with
bias-dependent differential capacitance $C_{i}(V_{i})$.

Take a SSS of Eq.(\ref{gwn02}) and (\ref{gwn03})
\begin{equation}
    V_{i}(t)\equiv V_{i}^{0}
\end{equation}
which should satisfy
\begin{equation}
    I_{i-1}(V_{i-1}^{0})=I_{i}(V_{i}^{0})
\end{equation}
and\vskip -0.8cm
\begin{equation}
    \sum_{i=1}^{N}V_{i}^{0}=U.
\end{equation}
Follow the standard analysis of linear stability analysis as WN
did\cite{wn}, we take a time-dependent perturbation of the
following form
\begin{equation}
    V_{i}(t)=V_{i}^{0}+A_{i}\exp(\lambda t).
\end{equation}
Substitute it into Eq.(\ref{gwn02}) and (\ref{gwn03}) and take the
linear terms of $\delta V_{i}(t)$, we have
\begin{eqnarray}
    C_{i}^{0}\lambda+G_{i}^{0}A_{i}=C_{i-1}^{0}\lambda+G_{i-1}^{0}A_{i-1}\nonumber\\
    \sum_{i=1}^{N}A_{i}=0
\label{sa01}
\end{eqnarray}
where
\begin{eqnarray}
    C_{i}^{0}\equiv C_{i}(V_{i}^{0}) \\
    G_{i}^{0}\equiv \frac{dI_{i}}{dV_{i}}\Big|_{V_{i}=V_{i}^{0}}
\end{eqnarray}
are the differential capacitance and differential conductance of
the $i-$th barrier at $V_{i}=V_{i}^{(0)}$. The solution of
Eq.(\ref{sa01}) has the following form
\begin{eqnarray}
    A_i=\frac{A}{C_{i}^{0}\lambda+G_{i}^{0}}
\end{eqnarray}
where $A$ is a non-zero constant. Substitute it back into
Eq.(\ref{sa01}), we find that a non-zero solution should satisfy
\begin{eqnarray}
    F(\lambda)=\sum_{i=1}^{N}\frac{1}{C_{i}^{0}\lambda+G_{i}^{0}}=
    \sum_{i=1}^{N}\frac{1/C_{i}^{0}}{\lambda+G_{i}^{0}/C_{i}^{0}}=0
\label{sa02}
\end{eqnarray}
The SSS is stable if and only if none of $\lambda_{i}$ satisfying
Eq.(\ref{sa01}) is positive.

\section{Effect of differential capacitance}
\label{NDC}

In this section, we shall discuss the effect of differential
capacitance on stability of the SSSs. We shall first consider the
case when all the differential capacitances are positive, i.e.,
$C_{i}^{0}>0$. As shown in Eq.(\ref{sa02}) in Sec.\ref{GWN}, an
SSS is stable if and only if none of $\lambda_{i}$ satisfying
$F(\lambda_i)=0$ is positive. In the case when $C_{i}^{0}>0$ for
all barriers, let us denote $\lambda_i=-G_{i}^{0}/C_{i}^{0}$ and
$\lambda_1\le\lambda_2\le\cdot\cdot\cdot\le\lambda_N$. WN found a
method to analyze the root of $F(\lambda)=0$\cite{wn}, and it is
easy to show that their method works when all $C_i^0$'s have the
same sign. Following WN's method\cite{wn} we obtain : (1)
$F(\lambda)=0$ has no negative root when $\lambda_i>0$ for all
barriers; (2) when $\lambda_1<0$ and $\lambda_2>0$, $F(\lambda)=0$
has one negative root if and only if $F(\lambda=0)>0$. Since
$C_i^0>0$ for all barriers, the above conclusions lead to the same
conclusions mentioned in Sec.\ref{WN} : (1) an SSS is stable if
all barriers have PDR, i.e., $G_{i}^{0}>0$; (2) if one and only
one barrier has NDR, then the SSS is stable when
\begin{eqnarray}
   F(\lambda=0)=\sum_{i=1}^{N}[G_{i}^{0}]^{-1}
   =\sum_{i=1}^{N}[I^{\prime}(V_{i}^{0})]^{-1}>0.
\label{edc01}
\end{eqnarray}
Therefore, the same proof as in Sec.\ref{WN} leads us to the
conclusion that in the generalized WN model of Eq.(\ref{gwn02})
and (\ref{gwn03}) there is always a stable SSS in the absence of
negative differential capacitance.
\begin{figure}[t]
\begin{center}
  \includegraphics[height=6cm,width=8cm]{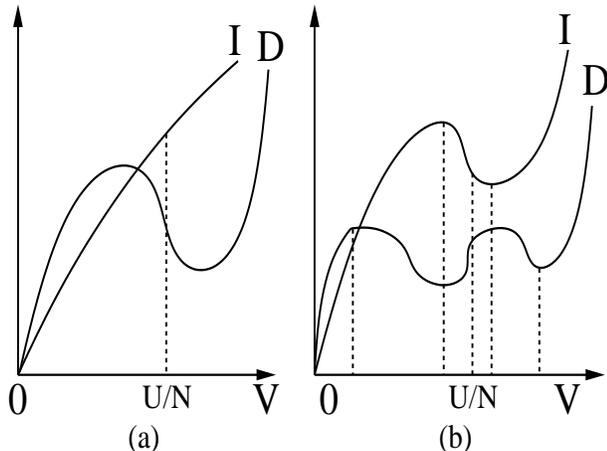}
\end{center}
\vskip -0.8cm \caption{Curves of $I$ and $D$ vs. $V$ for two
special cases which can lead to SSCO. (a) Barriers with negative
differential capacitance and with no NDR region; (b) barriers with
two bias regions of negative differential capacitance and with an
NDR region in between.} \label{ID_V} \vskip -0.8cm
\end{figure}

Now Let us see what can happen when negative differential
capacitance is introduced. It should be noted that since at
present we cannot have a reasonable way to expect {\it negative}
differential capacitance in the original superlattice system, the
model in this case can only be considered as a toy model. However,
this toy model may be a minimum model for the phenomenon of
self-sustained current oscillation. We shall show by a simple
example that negative differential capacitance can lead to a
non-stationary state. Suppose that all barriers have the same
$I-V$ curves with no NDR region, i.e., $G_{i}>0$, while $D-V$
curves are different and have a region of negative differential
capacitance as shown in Fig.\ref{ID_V}(a). It is obvious that in
this case the system has only one SSS at $V_i^0=U/N$. Let us
consider the case when $V_i^0=U/N$ drops into the negative
differential capacitance region of every barrier, i.e., $C_i^0<0$
for all barriers. Then, with WN's method it is easy to check that
this SSS is unstable if two of $\lambda_i=-G_{i}^{0}/C_{i}^{0}>0$
are not equal. Thus the system has no stable SSS and must be in an
SSCO state. This means that the generalized model with negative
differential capacitance and without NDR can lead to an SSCO
state. However, the model with only negative differential
capacitance is not enough because the $I-V$ oscillation behavior
of stable SSSs observed in experiments can only be included by
NDR. The SSCO behaviors can also occur when both negative
differential capacitance and NDR exist in the model. For example,
consider the case shown in Fig.\ref{ID_V}(b) where each barrier
has one NDR region between two negative differential capacitance
regions. In this case, the number of SSSs is more than one. One
can show that all SSSs are unstable when the average $V_i=U/N$
drops in the region of both PDR and negative differential
capacitance. Therefore, the generalized WN model with both NDR and
negative differential capacitance may be considered as a possible
minimum toy model for self-sustained current oscillation.

In a recent study, Sanchez et. al.\cite{bonilla1} have pointed out
that WN's original model will always tend to a stable stationary
state if it exists. However, their argument seems not fit for our
generalized model with negative differential capacitance.
Following their argument, the total current of each barrier in our
model is
\begin{equation}
    I_{total}=C_{i}(V_{i})\frac{dV_{i}}{dt}+I_{i}(V_{i})
\end{equation}
which leads to a series of equations for $V_i$
\begin{eqnarray}
    \frac{dV_i}{dt}=[C_{i}(V_{i})]^{-1}[I_{total}-I_{i}(V_{i})].
    \label{new00}
\end{eqnarray}
Put them into
\begin{eqnarray}
    \sum_{i}\frac{dV_i}{dt}=0,
\end{eqnarray}
we have
\begin{equation}
    I_{total}=\frac{\sum_{i}[C_{i}(V_{i})]^{-1}I_{i}(V_{i})}{\sum_{i}[C_{i}(V_{i})]^{-1}}.
    \label{new01}
\end{equation}
Put it back into Eq.(\ref{new00}), we have the equations for $V_i$
\begin{equation}
    \frac{dV_i}{dt}=\frac{1}{C_i(V_i)\sum_{j}[C_j(V_j)]^{-1}}\sum_{j}\frac{I_i(V_i)-I_j(V_j)}{C_j(V_j)}.
    \label{new02}
\end{equation}
Unlike the case in the Appendix of Sanchez et. al.\cite{bonilla1},
in the presence of negative differential capacitance $C_i<0$, the
total current $I_{total}$ in Eq.(\ref{new01}) can be
time-dependent and the equations of $V_i$'s in Eqs.(\ref{new02})
are coupled together through the factor $[C_j(V_j)]^{-1}$. Thus
the existence of limit circles and other types of SSCO behaviors
is possible in this toy model. Although no parameter in this toy
model represents directly for some external parameters in the
original SL system such as well doping, temperature and transverse
magnetic field, the presence of negative differential capacitance
and NDR may be viewed as the consequence of selecting appropriate
values of these external parameters in the original SL system.
Possible detailed relation between this toy model and the original
SL system needs further study and is not the aim of this paper.

In summary, our analysis shows that a generalized WN's model
without negative differential capacitance always has a stable SSS
and thus cannot have SSCO behaviors. This means that negative
differential capacitance is necessary for the existence of SSCO
behavior. We show by a simple example that the same model with
negative differential capacitance can have SSCO behavior. However,
the model with only negative differential capacitance is not
enough, since NDR is necessary for the $I-V$ oscillation behavior
of stable SSSs observed in experiments\cite{wn}. This means that
the generalized WN model with both negative differential
capacitance and NDR can serve as a possible minimum toy model for
the phenomena of both self-sustained current oscillation and $I-V$
oscillation behavior of stable SSSs.

GX acknowledges the support of CNSF under grant No. 10347101 and
the Grant from Beijing Normal University.

\end{document}